\documentclass[review]{elsarticle}

\usepackage{lineno,hyperref}
\usepackage{epstopdf}

\journal{Journal of \LaTeX\ Templates}

\begin{document}

\begin{frontmatter}

\title{Ultrawide color gamut single-pixel dynamic color manipulation based on yarn muscles–graphene MEMS}
\tnotetext[mytitlenote]{Fully documented templates are available in the elsarticle package on \href{http://www.ctan.org/tex-archive/macros/latex/contrib/elsarticle}{CTAN}.}

\author[mymainaddress,mysecondaryaddress]{Hongxu Li}
\author[mythirdaddress]{Bo Long}
\author[mysecondaryaddress]{Tao Wang}
\author[mysecondaryaddress]{Feng Zhou}
\author[mymainaddress]{Zhengping Zhang\corref{mycorrespondingauthor}}
\cortext[mycorrespondingauthor]{Corresponding author}
\ead{zpzhang@gzu.edu.cn}

\address[mymainaddress]{College of Big Data and Information Engineering, Guizhou University, Guiyang 550025, P.R.China}
\address[mysecondaryaddress]{Guizhou Light Industry Technical College, Guiyang 550025, P.R.China}
\address[mythirdaddress]{Institute for Metrology and Calibration of Guizhou, Guiyang 550003, P.R.China}



\begin{abstract}
This work investigated the single-pixel color modulation in a composite structure of yarn muscles–graphene mechanical system and photonic crystal multimode microcavity. The position of graphene in the microcavity is modified by changing the yarn muscles stretching using different current levels. This helps in adjusting the light absorption of graphene to different colors. Hence, red, green, blue, and their mixed colors can be displayed using a single pixel; color gamut of this system can reach 96.5\% of RGB. The proposed system can avoid the spontaneous oscillation caused by large strain energy. This solution can provide insights into the design of low-power, ultrahigh-resolution, and ultrawide color gamut interferometric modulator display technologies.
\end{abstract}


\end{frontmatter}


\section{Introduction}
With the rapid development of multimedia technology, an increasing number of emerging fields, such as virtual reality and wearable displays, have demanded high requirements for display technology, including high resolution, wide color gamut, and low-power consumption. The traditional display screen comprises three sub-pixels of red, green, and blue, which create obstacles to further improve the display’s high resolution \cite{49kim2012high,50kim2011unusual}.

The display of red, green and blue, and their mixed colors in a single pixel can greatly improve the display resolution and has thus become, the focus of this research. Chun $et$ $al$. stacked two different colors of light-emitting diodes (LEDs) vertically to obtain LED that can emit blue, yellow, and their mixed colors \cite{47chun2014vertically}. Kang $et$ $al$. realized a vertically stacked subpixel-type array structure in addition to a horizontally aligned subpixel-type structure, which provides a threefold increase in display resolution over the horizontally aligned subpixel-type \cite{53kang2017monolithic}. Similarly, Kim $et$ $al$. achieved the display of red, green, blue, and their mixed colors by vertically stacking red, green, and blue LEDs in a single pixel of 4 $\mu$m such that their array density reaches 5,100 pixels/inch \cite{48shin2023vertical}.

However, the current research on single-pixel panchromatic displays is mainly focused on active devices, such as LEDs, with less focus on passive devices. Recently, Xu $et$ $al$. reported that the position of graphene in the microcavity can be changed by adjusting the voltage to change the deformation of graphene, thus realizing the display of monochrome light or multiple mixed colors within a single pixel \cite{33xu2021single}. To achieve this, they combined graphene micro-electrical–mechanical systems (MEMS) with three resonant standing wave mode photonic crystal microcavities with corresponding wavelengths of red, green, and blue primary colors. Compared with conventional MEMS \cite{1chan2016continuous,51han2022mems,52ishizuka2015mems}, graphene has superior mechanical and optoelectronic properties, higher Young’s modulus, larger specific surface area, very low surface density, and excellent ductility \cite{2liu2007ab,3wang2020circular,4shen2022quantized,5deng2022graphene,41deop2022optical,42novitsky2022random,44nag2022graphene,45esfandiari2022recent}. Hence, electro–optical modulators of graphene MEMS have been successfully applied in optical projectors and simplified spectrometers \cite{8reserbat2016electromechanical}. For instance, Dejan Davidovikj $et$ $al$. utilized a phase-sensitive interferometer to visualize the vibrations of micrometer-scale graphene nanodrums at extremely high frequencies \cite{54davidovikj2016visualizing}. Meanwhile, Houri $et$ $al$. discovered that the bilayer graphene  membrane can reproduce the interference effects of an interferometric modulator display. Under white light illumination, the downward-bending bilayer graphene membrane displayed Newton's rings phenomenon at an approximately 1 bar pressure difference between the inside and outside of the cavity \cite{55cartamil2016colorimetry}. Furthermore, they studied the electro-optical response of bilayer graphene pixels and successfully manufactured a reflective graphene-based display with a high-resolution image (2500 pixels per inch) \cite{56cartamil2018graphene}. This experiment underscored the excellent color reproducibility of graphene films under electro-optical modulation, making them an ideal choice for high-resolution display pixels. Displays based on graphene mechanical pixels showcase undeniable advantages of being ultra-thin and high-resolution, as well as being more durable, energy-efficient, flexible, and easy to control compared to LED screens. However, the current graphene MEMS are mainly based on graphene elastic deformation \cite{6fan2019graphene,7fan2019suspended,8reserbat2016electromechanical}, and the spontaneous oscillations due to large strain energy cause graphene MEMS to achieve excellent performance MEMS oscillators \cite{9ye2018glowing,11arjmandi2017large,12kim2018nano} rather than light modulators. 

Here, we introduced yarn muscles into the design to further reduce the large deformation energy caused by the elastic deformation of graphene, reduce the modulation voltage and energy consumption, and expand its color gamut. Yarn muscles stretch, contract, twist, rotate, or bend by converting the input electrical, thermal, or chemical energy into mechanical energy under external heat, light, or electrical stimulation \cite{13xu2019artificial,14aziz2020artificial,15foroughi2019carbon,16chu2021unipolar,46oveissi2021tough}. Owing to its small size, high freedom of motion, good environmental adaptability, and absence of large deformation energy and spontaneous vibration, yarn muscles have broad application prospects in robotics, flexible mechanical electronics, biomedicine, and precision minimally invasive surgery \cite{17zhu2021two,18cheng2021simultaneous,19hu2022fast,20li2021gel}. These devices, based on yarn muscles, are actively optically controllable. They have the remarkable ability to lift objects that weigh more than 650 times their own weight and can withstand a strain of up to 1000\%. Additionally, they demonstrate long-term elasticity and resilience even after undergoing 105 cycles of deformation \cite{57kanik2019strain}.

Specifically, photonic crystal microcavities were designed with three resonant standing wave modes corresponding to red, green, and blue primary wavelengths. When the yarn muscles pull the graphene toward a stronger light field, the absorption is high and transmission is low, while this reverses when the yarn muscles pull the graphene toward a weaker light field, i.e., now the absorption is low and transmission is high. Since different colors of light have different wavelengths, they have different field distribution in the microcavity. Therefore, to change the position of graphene in the microcavity, the absorption of graphene to different colors of light is first adjusted, and then, the output of monochromatic light or multiple mixed colors of light within a single pixel is realized. This research proposal can provide guidance and new ideas for designing and fabricating low-power, ultrahigh-resolution, ultrawide color gamut interference modulator display technology.

\section{Theoretical analysis}

\begin{figure}
	\centering
	\includegraphics[width=\linewidth]{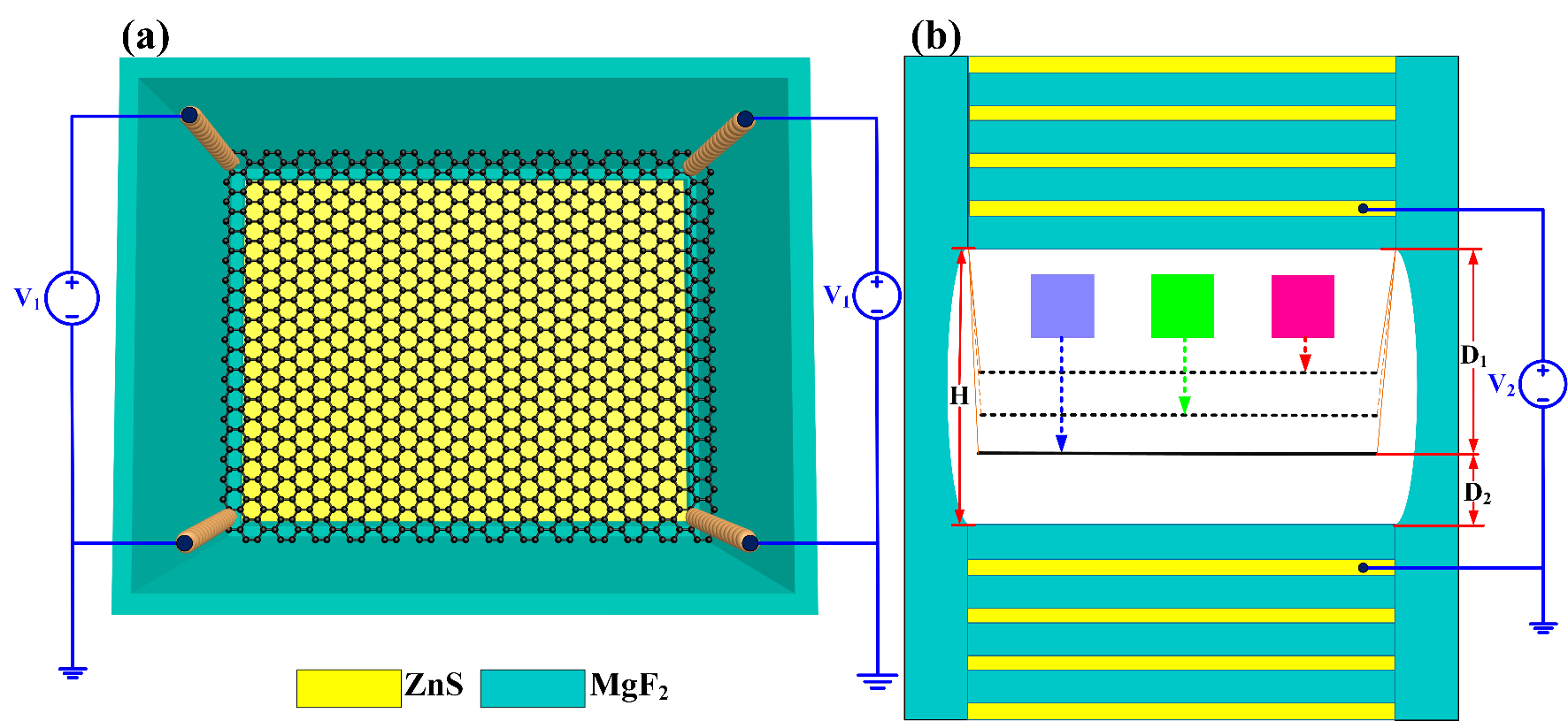}
	\caption{(a) Top cross-sectional view of the optical microcavity with graphene; (b) side view of the designed structure (inset shows the modulation colors when graphene is in different positions).}
	\label{Fig1}
\end{figure}

A diagram of the stereo results of the studied photonic crystal microcavity is shown in Fig. \ref{Fig1}. The microcavity is flanked by Bragg mirrors composed of two materials, ZnS and MgF$_2$, with alternating refractive indices ($n_1$ and $n_2$) and are denoted in yellow and blue, respectively, and the refractive indices of the dielectric layers are taken to be $n_1$ = 1.38 and  $n_2$ = 2.5, respectively, in the calculation, with all layers being nonmagnetic ($\mu$ = 1). The thicknesses of the relevant layers are $d_1$ = $\lambda_0/4n_1$ and  $d_2$ = $\lambda_0/4n_2$ , respectively, and $\lambda_0$ is the central wavelength, which is taken as 530 nm here. the Bragg mirror period is M and is taken as M = 4 in the calculations. Similar Bragg mirror structures can be easily fabricated by multilayer deposition \cite{21hillmer2002potential,22irmer2003ultralow,23prott2003modeling,24irmer2005surface}, which provides a reasonable guarantee for the experimental realization of our scheme.

Intriguingly, the position of a graphene film can be precisely adjusted by the stretching of yarn muscles, which can be controlled by applying an electric current. It is worth mentioning that muscle fibers can be coupled with graphene through the van der Waals force between the electrodes and graphene. This coupling mechanism allows for their interaction and integration. Furthermore, by applying a voltage V$_2$ between the upper and lower Bragg gratings, the conductive layer and graphene can be made to mutually attract or repel; this helps to effectively avert undesired graphene vibrations inside the cavity. The refractive index of a single layer of graphene is $n_g$ = 3.0+$i(C_1/3)\lambda$, where $C_1$ = 5.446 $\mu m^{-1}$, and $\lambda$ is the wavelength \cite{25jiang2018broad,26bruna2009optical}, and the thickness of a monolayer of graphene is 0.34 nm. The thickness of the air layer of graphene with the Bragg mirror above, D$_1$, is 1030 nm and that with the Bragg mirror below, D$_2$, is 49 nm, as shown in Fig. \ref{Fig1}(a). The transmission matrix method is often used for the analysis of photonic crystal microcavities composed of Bragg mirrors.

We analyzed the transport of light in a Bragg grating-based photonic crystal microcavity using the transfer matrix method. The incident light is separated into two modes: transverse magnetic($TM$), where the magnetic field is parallel to the interface, and transverse electric($TE$), where the electric field is parallel to the interface. For $l$-$th$ layer, the electric (magnetic) field of the $TE$ mode ($TM$ mode) beam $E_l^{TE}(z,x)$ ($H_l^{TM}(z,x)$) is given:

\begin{equation}
	\left\{
	\begin{array}{rcl}
		E_l^{TE}(z,x)=[A^E_le^{-k_{liz}(z-z_l)}e^{ik_{lrz}(z-z_l)}+B^E_le^{k_{liz}(z-z_l)}e^{-ik_{lrz}(z-z_l)}]e^{-k_{liz}x}e^{ik_{lrz}x}
		\\H_l^{TM}(z,x)=[A^H_le^{-k_{liz}(z-z_l)}e^{ik_{lrz}(z-z_l)}+B^H_le^{k_{liz}(z-z_l)}e^{-ik_{lrz}(z-z_l)}]e^{-k_{liz}x}e^{ik_{lrz}x}\\
	\end{array}, \right.\label{eq1}\\ 
\end{equation}

Where $k_l$=$k_{lr}$+$i$$k_{li}$ is the wave vector of light and  $z_l$ is the position of the $l$-$th$ layer in the $z$-direction. The electric and magnetic field components of the light at the boundary of each dielectric layer must satisfy the boundary conditions by which the electromagnetic fields are connected between two adjacent layers.

Thus, the relationship between the electric field amplitude of the ($l$+1)-$th$ layer of the microcavity and incident light irradiating the first layer is expressed in equation (2),

\begin{equation}
	\left(
		\begin{array}{ccc}
			A^{E,H}_{l+1}\\B^{E,H}_{l+1}
		\end{array} 
	\right) = T^{l+1\leftarrow l}T^{l\leftarrow l-1}...T^{2\leftarrow 1}
	\left(
		\begin{array}{ccc}
			A^{E,H}_1\\B^{E,H}_1 
		\end{array} 
	\right)  \\=
	\left(
		\begin{array}{ccc}
			T^{E,H}_{11}\quad T^{E,H}_{12}\\T^{E,H}_{21}\quad T^{E,H}_{22} 
		\end{array} 
	\right) 
	\left(
		\begin{array}{ccc}
			A^{E,H}_1\\B^{E,H}_1
		\end{array} 
	\right),
\end{equation}%

Since no light waves are reflected once the light passes through an optical microcavity, the electromagnetic field in the ($l$+1)-$th$ layer has no light waves propagating in the negative direction of the $z$-axis, i.e.,  $B^{E,H}_{l+1}$ =  0. Therefore, the microcavity transmittance is obtained as \cite{27zhou2021ultrawide,28liu2019two},

\begin{equation}
		t=\frac{|A_{l+1}|^2}{|A_1|^2}=\left|T_{11}-\frac{T_{12}T_{21}}{T_{22}}\right|^2\\ 
\label{eq3}\\ 
\end{equation}

\section{Numerical result analysis}

\begin{figure}
	\centering
	\includegraphics[width=\linewidth]{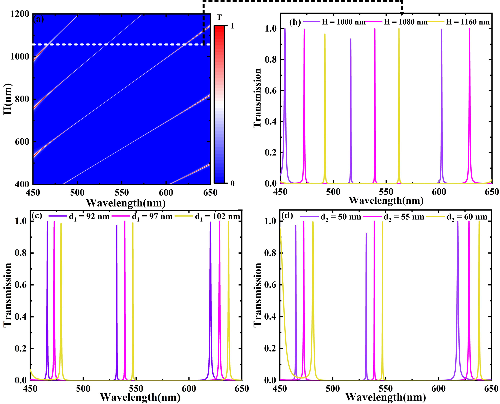}
	\caption{(a) Variation in the transmittance spectra of a photonic crystal microcavity with defect layer thickness (microcavity height); (b) Transmission and reflection spectra of the microcavity when the cavity height is at 1000, 1080, and 1160 nm; (c, d) Variation of the microcavity transmittance spectra with d$_1$ and d$_2$.}
	\label{Fig2}
\end{figure}

The resonant state of the microcavity is sensitive to the thickness of the intermediate defect layer, and the resonant wavelength $\lambda_c$ satisfies the condition $m_i\lambda_c/2\propto L_c$, where $L_c$ is the optical range of the microcavity, $\lambda_c$,  is the resonant wavelength, and $m_i$ is a positive integer \cite{29jiang2018broad}.  As shown in Fig. \ref{Fig2}(a), we conducted a comprehensive scan of the air cavity thickness in the middle of the defect layer and detected multiple modes at 400 nm to 1200 nm. We then selected the thickness indicated by the white line (H = 1080 nm) to conform to the wavelengths of the primary colors, i.e., red, green, and blue. The transmittance in the immediate vicinity of this thickness are depicted in Fig. \ref{Fig2}(b), which highlights zero absorption, i.e., no loss in the microcavity. We also performed an in-depth study on the influence of media layer thicknesses that constitute the Bragg grating on the microcavity light transmission spectra. Remarkably, as d$_1$ and d$_2$, i.e., the thicknesses of MgF$_2$ and ZnS, respectively, increase, the photonic crystal transmission spectra undergoes a discernible red shift, as displayed in Figs. \ref{Fig2}(c-d).

\begin{figure}
	\centering
	\includegraphics[width=\linewidth]{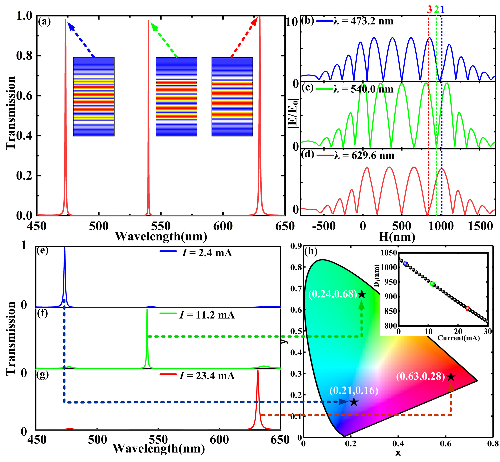}
	\caption{(a) Transmission spectrum of the graphene-free optical microcavity; (b-d) shows the electric field distribution at wavelengths of 473.2, 540.0, and 629.6 nm, respectively; (e-g) Transmission spectra corresponding to currents $I$ of 2.4, 11.2, and 23.4 mA, respectively. (h) Color coordinates of 2.4, 11.2, and 23.4 mA and their corresponding positions in the 1931 CIE chromaticity diagram. The illustration depicts the relationship between yarn muscles and electric current during current modulation.}
	\label{Fig3}
\end{figure}

Within the microcavity, light undergoes numerous reflections between two Bragg gratings positioned on either sides of the cavity. These reflections generate a standing wave distribution that arises through superposition coupling. In the absence of graphene photonic crystals, the microcavity can accommodate three resonant modes, whose wavelengths align with the primary colors of red, green, and blue. These resonant modes emerge when the microcavity height is fixed at approximately 1080 nm and local optical fields are created at various positions throughout the microcavity, resulting in three resonant transmission peaks in the spectral range. The transmittance rates of these peaks are impressive, with values of 98.0\%, 97.6\%, and 99.9\% for the red, green, and blue modes, respectively. As illustrated in Fig. \ref{Fig3}(a), these peaks can generate red, green, and blue colors. Moreover, the narrow resonances, paired with high Q values, guarantee the production of exceptionally pure colors.

The absorption of light by graphene is intricately linked to the field strength and distribution of its structure. As demonstrated in Figs. \ref{Fig3}(b-d), the electric field distribution in all three modes exhibits heightened electric field strength. Here, E and E$_0$ refer to the electric field amplitude at the resonant wavelength and electric field amplitude of the incident light, respectively \cite{30song2019colors,31he2020dynamically}. The incident light generates three sets of standing waves by reflecting back and forth between the microcavity’s photonic crystal mirrors, as detailed in Fig. \ref{Fig3}(a).

The absorption of light of different colors can be modified by manipulating yarn muscles to adjust the position of five-layer graphene in a photonic crystal microcavity. Specifically, applied electric currents can cause reversible strain in the yarn muscles. The interplay between the current and yarn muscles can be described using the equation D$_1$ = P$_1$ $\times$ $I^2$ + P$_2$ $\times$ $I$ + P$_3$, where D$_1$ denotes the length of the yarn muscle, $I$ represents the current magnitude, and P$_1$ = 0.0219, P$_2$ = - 7.8571, and P$_3$ = 1030, as reported in reference \cite{32tu2021biomimetic}. A graphical presentation of this relationship is depicted in the inset of Fig. \ref{Fig3}(h).

The equilibrium position of graphene in a microcavity controls the transmission characteristics of its three modes. The is because when graphene is located in the standing wave belly part of the microcavity (where the optical field is strong), graphene light absorption is high and light transmission is low; conversely, when graphene is located in the standing wave node to the microcavity (where the optical field is weak), graphene light absorption is low and transmission is high. The intensity of transmitted light displays positional dependence when graphene is situated between the nodes and antinodes of a standing wave within the cavity. As the three resonant modes have different field distributions, graphene can absorb different colors of light by adjusting the position of graphene in the microcavity, allowing red, green, and blue colors to pass through individually or as a mixture. In summary, the device can adjust the position of the graphene film in the cavity by applying the current to control the stretching of the yarn muscles \cite{32tu2021biomimetic}, thereby allowing the current to modulate the intensity and color of the three modes of transmission.

Thus we have discussed three typical positions (1012, 944, and 858 nm) of graphene in the microcavity, corresponding to the dashed positions 1–3 in Figs. \ref{Fig3}(b-d) and their corresponding transmission spectra are shown in Figs. \ref{Fig3}(e-g). When the yarn muscles of initial length 1012 nm are applied with a current of 2.4 mA, the graphene equilibrium position is pulled to 1012 nm since the graphene is located in the wave node of the B-mode standing wave. Furthermore, the graphene absorbs low B-mode light and has a low effect on the B-mode electric field distribution, while it can also absorb more G and R-mode standing wave; thus, graphene film allows the B-mode to pass through. Its transmission is also at its peak at this time and the interference color is blue, as illustrated in Fig. \ref{Fig3}(e). Similarly, when the applied current is 11.2 mA (23.4 mA), the length of the yarn muscle shrinks to 944 nm (858 nm) and the graphene is located in the G (R) mode standing wave node. Now, the graphene film absorbs low light of the G (R) mode, has low effect on its electric field distribution, and the interference color appears green (red), as shown in Figs. \ref{Fig3}(f, g). When the transmission amplitude of the B (G, R) mode resonance wavelength reaches its maximum, the remaining two are close to zero, forming a color that is not monochromatic; however, as a result of multiple color integrals acting together, the color at these points is relatively pure.  The above three points, as shown in Fig. \ref{Fig3}(h), are located at (0.21, 0.16), (0.24, 0.68), and (0.63, 0.28), respectively, and this results fully demonstrate  the ultrawide color gamut characteristic of single-pixel color modulation.

\begin{figure}[ht]
	\centering
	\includegraphics[width=\linewidth]{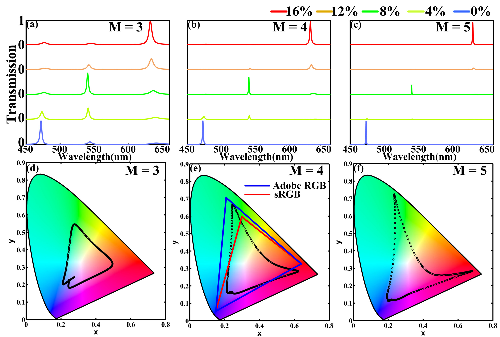}
	\caption{(a-c) Transmission spectra changes under different strains applied to yarn muscles for Bragg grating period M ranging from 3 to 5. (d-f) Corresponding changes in modulated colors as the Bragg grating period varies.}
	\label{Fig4}
\end{figure}

In our subsequent analysis, we conducted calculations on the transmission spectra and color gamut of graphene-based Bragg gratings with varying periods (M = 3, M = 4, and M = 5) under TM-polarized light incidence. As depicted in Figs. \ref{Fig4} (a-c), the transmission spectra of this structure exhibit a progressive modulation of RGB peak values when the yarn muscles undergo contraction, transitioning from a state of relaxation (0\%) to 16\%. Remarkably, a clear trend is observed as the number of Bragg grating periods contributing to the formation of photonic crystal microcavities increases from left to right: the width of the highest transmission peak narrows and the corresponding Q value shows an increasing trend. This behavior arises from the dependence of bandgap width and quality factor on the number of Bragg grating periods, ultimately influencing the color gamut peak width of the transmitted spectra and consequently affecting its chromatic properties. Furthermore, our investigation in Figs. \ref{Fig4} (d-f) reveals a conspicuous association between the number of Bragg grating periods and the expansion of the transmission color gamut. Additionally, a comparative analysis between M = 4 and M = 5 highlights that while the color gamut expands for M = 5, there is a notable decrement in color mixing, particularly in the red-green combination. Hence, based on these findings, we judiciously select M = 4 as the optimal microcavity structure design in our study.

\begin{figure}[ht]
	\centering
	\includegraphics[width=\linewidth]{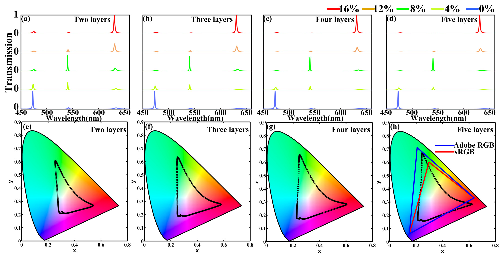}
	\caption{(a-d) Changes in transmission spectra as the number of graphene layers varies under different strains applied to yarn muscles. (e-f) Corresponding changes in modulated colors as the number of graphene layers varies.}
	\label{Fig5}
\end{figure}

Furthermore, we conducted an analysis to explore the influence of different numbers of graphene layers, with a fixed period of M = 4, on the transmission spectra and color gamut under TM-polarized light incidence. As depicted in Figs. \ref{Fig5} (a-d), the transmission spectra of this intricate structure exhibit a remarkable continuous modulation of RGB peak values as the muscular fibers undergo varying levels of contraction. This phenomenon can be attributed to the distinct spatial locations of the graphene layers within the photonic crystal microcavity, dictated by the dynamic stretching of the muscular fibers. Consequently, the absorption characteristics of graphene are inherently altered.	Investigating the spectra from left to right, intriguing observations emerge. While the impact on the highest peak remains negligible with an increase in the number of graphene layers, owing to their positioning in regions of lower optical intensity, the excitation of spectral modes corresponding to areas of stronger optical intensity showcases a more pronounced influence. The progressive increment in graphene layers accentuates absorption, eventually leading to a decline in the overall peak values. For instance, as illustrated in the first row of the transmitted spectra from left to right, the R-mode remains relatively unaffected, whereas discernible decrements are observed in the B-mode and G-mode, ultimately approaching near-zero values when the graphene layers reach a count of five. Intriguingly, the augmentation of graphene layers presents an opportunity for expanding the color gamut. This phenomenon is intricately linked to the amplified absorbance of graphene, ultimately enhancing the modulation amplitudes. A graphical depiction in Figs. \ref{Fig5} (e-h) elegantly demonstrates that thicker films engender a more vibrant manifestation of chromatic responses.

\begin{figure}[ht]
	\centering
	\includegraphics[width=\linewidth]{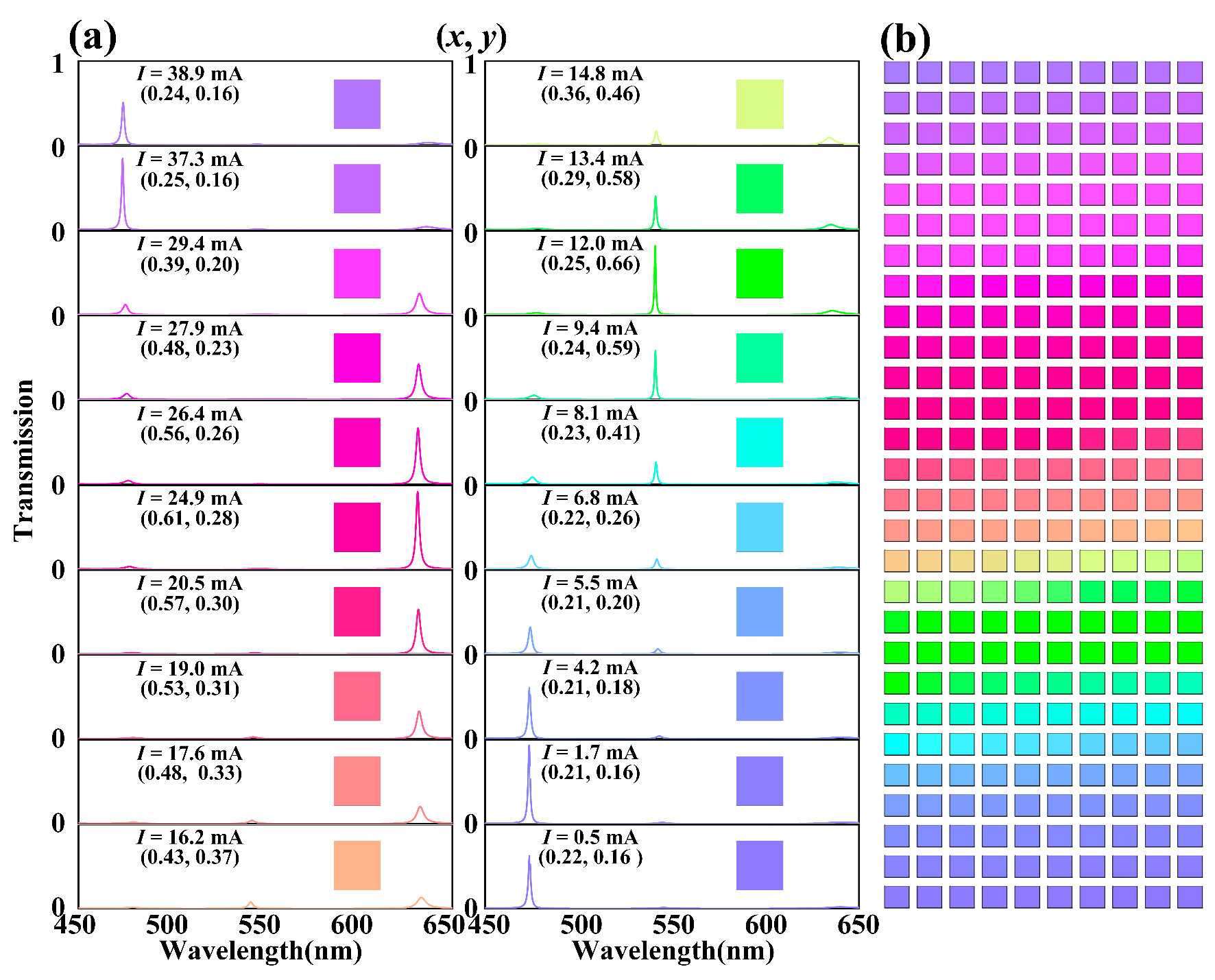}
	\caption{(a) Simulated transmission spectra of graphene position increasing from 752 nm to 1031 nm in an optical microcavity; (b) D$_1$ increases from 752 nm to 1031 nm with a 1 nm-gradient in the palette.}
	\label{Fig6}
\end{figure}

In summary, we conducted an in-depth analysis of the RGB channels of graphene micromechanical pixels across diverse layers and periods. We obtained valuable insights into the color range of graphene single pixels and their diverse variations with currents . As depicted in Figs. \ref{Fig4} and \ref{Fig5}, the increase in the balance position of graphene deflection presents the black point trajectory, revealing the shift in the equilibrium position of graphene, which highlights the color trend. This indicates a positive realization of electro–optical modulation between the three primary colors within a single pixel. Impressively, as shown in Fig. \ref{Fig5}(h), the five-layer graphene mechanical system demonstrated color gamut coverage of approximately 96.5\% of standard RGB (sRGB) and 71.1\% of Adobe RGB. The dynamic modulation color gamut of the graphene MEMS single pixels matched impeccably with standard RGB color. Importantly, our findings surpass the outcomes of previous works \cite{33xu2021single} by 1.3 times, highlighting the novelty and utility of this exploratory research.

Remarkably, the system demonstrated extensive visible light dynamic modulation without the need for complex structures. Furthermore, we incorporated a TM-polarized light incidence scenario, utilizing a Bragg grating structure with a periodicity of four and five graphene layers. To investigate the impact of current modulation, we systematically varied the initial equilibrium position of graphene from 752 nm to 1031 nm in our design. As shown in Fig. \ref{Fig6}(a), diverse visible light transmission spectra and their corresponding CIE chromaticity coordinates ($x$, $y$) and color blocks were obtained across different current modulations. Moreover, with the gradual increase in D$_1$, the interference colors corresponding to the transmission spectra exhibited a cyclic pattern, covering the entire visible light range before finally returning to blue. Fig. \ref{Fig6}(b) showcases the joint color blocks of the transmission colors with a step of 1 nm, vividly illustrating the brilliant rainbow colors achievable with a single pixel, thereby highlighting the immense potential of the graphene MEMS system in emerging display technologies \cite{34sun2017all,35sun2018real,36kim2019generation}.

It is worth concluding by highlighting the remarkable experimental feasibility underlying the findings presented in this study. While simulations form the basis of our results, it is crucial to emphasize that all the parameters employed were carefully selected in accordance with established experimental data. Particularly noteworthy is the fabrication process of the graphene-dielectric distributed Bragg reflector (DBR) composite structure captured in Fig. \ref{Fig1}, as it stems from well-documented methodologies in prior research. By harnessing plasma-enhanced chemical vapor deposition (PECVD) techniques, the dielectric DBR can be deposited onto the epitaxial structure at reduced temperatures \cite{21hillmer2002potential}. Moreover, the integration of the DBR with a sacrificial polymer layer offers a pathway for creating cavities, as demonstrated by previous studies \cite{23prott2003modeling}. Additionally, the growth of multi-layer graphene films via chemical vapor deposition (CVD) can be followed by their subsequent introduction into the DBR cavity using a semi-dry transfer technique \cite{39lin2019air}. Notably, the simplicity and cost-effectiveness associated with the growth of one-dimensional films, combined with relatively straightforward manufacturing processes, pave the way for practical production and enhanced scalability.

\section{Conclusions}
In this study, we investigated the color modulation design for single pixels in yarn muscle–graphene MEMS and photonic crystal multimode microcavity composite structures. Specifically, we designed a photonic crystal microcavity that produces resonance modes for red, green, and blue wavelengths, generating local optical fields at different positions within the cavity. Interestingly, graphene demonstrated high optical absorption and low transmission in strong optical fields, while the opposite trend was observed in weak optical fields. By utilizing electric currents to adjust the length of the yarn muscles, we could effectively regulate the position of graphene within the microcavity and achieve the display of primary and mixed colors of red, green, and blue. Moreover, the system’s color gamut range achieved an impressive 96.5\% of sRGB, while the graphene MEMS prevented spontaneous oscillations due to large strains, thereby overcoming a significant limitation of traditional graphene MEMS. Overall, our findings provide crucial insights and suggestions for designing and producing interference modulator displays with ultrahigh resolution and wide color gamut range.

\section*{Acknowledgements}
This work was supported by the National Natural Science Foundation of China (NSFC) (Grant no. 62364004, 62174040), the Young scientific and technological talents growth project of the Department of Education in Guizhou Province (Grant no. KY[2022]184), Nature Science Foundation of Guizhou Minzu University ([2018]5773-YB02, GZMUZK[2021]YB07), and the Guizhou Provincial Science and Technology Projects (Grant no. [2020]1Y026, ZK[2022]211).

\section*{References}

\bibliography{mybibfile}

\end{document}